\documentclass[useAMS,usenatbib,letters]{mn2e}
\usepackage{aas_macros,graphicx,times,multirow}
\usepackage[]{color}
\title[Continuum radio emission associated with Geminga]
  {Detection of continuum radio emission associated with Geminga}
\author[A.~Pellizzoni et al.]
{A.~Pellizzoni,$^{1}$\thanks{E-mail: apellizz@oa-cagliari.inaf.it} F.~Govoni,$^{1}$ P.~Esposito,$^{1}$ M.~Murgia$^{1}$ and A. Possenti$^{1}$\smallskip\\
$^1$INAF -- Osservatorio Astronomico di Cagliari, localit\`a Poggio dei Pini, strada 54, I-09012 Capoterra, Italy}
\date{Accepted 2011 June 14. Received 2011 June 13; in original form 2011 February 14}
\pagerange{\pageref{firstpage}--\pageref{lastpage}} \pubyear{2011}

\def\LaTeX{L\kern-.36em\raise.3ex\hbox{a}\kern-.15em
    T\kern-.1667em\lower.7ex\hbox{E}\kern-.125emX}
 
\begin{document}

\label{firstpage}
\maketitle

\begin{abstract}
A deep Very Large Array observation of the Geminga pulsar field led to the discovery, at a higher than 10$\sigma$ significance level, of radio emission trailing the neutron star proper motion. This $\sim$$10\arcsec$-long radio feature, detected with a flux of $\sim$0.4 mJy at 4.8 GHz, is marginally displaced ($2.7\pm1.8$ arcsec) from the pulsar (which, at any rate, is unlikely to contribute with magnetospheric pulsed emission more than 15\% to the total observed radio luminosity, $\sim$$10^{26}$ erg s$^{-1}$) and positionally coincident with the X-ray axial tail recently discovered by \emph{Chandra} and ascribed to the pulsar wind nebula (PWN). Overall, the Geminga radio tail is compatible with the scenario of a synchrotron-emitting PWN, but the present data do not allow us to discriminate between different (and not always necessarily mutually exclusive) possible processes for producing that. If this radio feature does not result from intrinsic peculiarities of Geminga, but its proximity and radio-quiet nature (both helping in not hindering the faint diffuse radio emission), other relatively near and energetic radio-quiet pulsars could show similar structures in dedicated interferometric observations.
\end{abstract}


\section{Introduction}

Geminga (PSR\,J0633+1746) is a nearby radio-quiet pulsar,\footnote{We refer to a pulsar as radio-quiet when, even though it may emit radio pulses, these cannot be detected at Earth.} discovered as a gamma-ray source and later identified as a neutron star (NS) through optical and X-ray observations (see \citealt{bignami96} for review). The pulsar period ($P\sim$237 ms) and its derivative ($\dot{P}\sim$$1.1\times10^{-14}$ s s$^{-1}$) correspond\footnote{See e.g. \citealt{lk05} for the formulae used to derive $\tau,~\dot{E}_{\mathrm{rot}}$ and $B_{\rm surf}$ from $P$ and $\dot{P}$.} to a spin-down age $\tau\sim$340 kyr, a spin-down power $\dot{E}_{\mathrm{rot}}=3.3\times10^{34}$ erg s$^{-1}$ and a surface magnetic field $B_{\rm surf}\sim 1.6\times 10^{12}$ G. \emph{Hubble Space Telescope} parallax measurements confirmed the proximity of Geminga ($d=250^{+120}_{-62}$ pc) and a proper motion corresponding to a transverse velocity of $\sim$210 km s$^{-1}$ \citep{faherty07}, exceeding the typical sound speed in the interstellar medium (10--30 km s$^{-1}$). Together with Crab and Vela, Geminga is one of known pulsars with the highest spin-down flux ($\dot{E}_{\mathrm{rot}}d^{-2}$), allowing detailed studies  of weak structures in the immediate vicinity of the NS and its interactions with the local interstellar medium.

Pulsar wind nebulae (PWNe) are common diffuse features surrounding NSs with very different spin-down ages, that result from the interaction of the  relativistic pulsar wind and the ambient medium, producing shocks and outflows that can be observed, mostly because of synchrotron and inverse Compton (IC) radiation, in a very broad range from radio to gamma-ray energies (see \citealt{kaspi06,gaensler06,kargaltsev08,pellizzoni10short}). 

The complex PWN structures mostly depend on the particular wind outflow geometry and on the ratio of the pulsar speed to the sound speed in the ambient medium \citep{bucciantini05,romanova05,bogovalov05}. Well-resolved  equatorial ``tori" and axial (with respect to the spin axis) ``jets" are seen around pulsars moving in the interstellar medium with subsonic velocities (e.g. Crab and Vela pulsars; \citealt{whj00,helfand01}), while in pulsar moving with supersonic speed, as is the case of Geminga, bow-like shapes ahead of the pulsar and tails behind are typically observed \citep{pellizzoni05}.

In fact, \emph{XMM-Newton} observations of Geminga in April 2002 revealed two $\sim$2$\arcmin$-long tails behind the pulsar, approximately symmetric with respect to the sky projection of the pulsar's trajectory \citep{caraveo03}. Geminga was then observed by \emph{Chandra} in February 2004 unveiling a new structure $\sim$25$\arcsec$-long and $\sim$5$\arcsec$-thick, with a surface brightness $\sim$40 times higher than that of the \emph{XMM-Newton} tails, starting at the pulsar position and aligned with the proper motion direction \citep{dcm06}. A faint arc-like structure was also reported 5--7 arcsec ahead of the pulsar \citep{pavlov06}. A deeper \emph{Chandra} observation carried out in August 2007 confirmed the existence of the axial tail (50$\arcsec$-long) and of the two outer tails discovered by \emph{XMM-Newton} \citep{pavlov10} providing an overall X-ray PWN luminosity of $L_{0.3-8~\mathrm{keV}}=3\times10^{29}d^2_{250}$ erg s$^{-1}$ ($\simeq$$10^{-5}d^2_{250}\dot{E}_{\mathrm{rot}}$). Comparing 2004 and 2007 \emph{Chandra} observations, \citet{pavlov10} found indication that the X-ray tail is composed of variable $\sim$5$\arcsec$-long sub-structures (streaming ``blobs"). This complex morphology was interpreted as synchrotron radiation possibly arising from the shocked pulsar wind collimated by the ram pressure and/or jets emanating along the pulsar's spin axis.

Apart from diffuse TeV emission around Geminga on larger scales ($>$$2\degr$) with respect to the X-ray nebula \citep{abdo09} and a possible detection in the mid-infrared \citep{danilenko11}, the Geminga PWN has not been detected in other energy bands so far. In particular, since the discovery of Geminga as a gamma-ray source, many observers have attempted to detect it at radio frequencies, both as a continuum and as a pulsating source, providing only tentative detections of weak pulsed emission at low frequencies ($\sim$100 MHz; see \citealt{kuzmin97,kl97,malofeev97,shitov97,shitov98,vats97,vats99}). \citet{kassim99} reported upper limits on the continuum emission (Very Large Array [VLA] data) at 74 MHz ($<$56 mJy) and 326 MHz ($<$5 mJy) reconciling their nondetection with previous pulsed detections by invoking intrinsic or extrinsic (refractive interstellar scintillation) variability. At higher frequencies, \citet{spoelstra84} reported an upper limit of 0.5 mJy at 21 cm (Westerbork Synthesis Radio Telescope data) and 1 mJy at 6 cm (VLA data). More recently, \citet{giacani05} presented a high-resolution (24$\arcsec$) study of the H\textsc{i} interstellar gas distribution around Geminga including a 1.4 GHz upper limit of $\sim$0.4 mJy in the pulsar direction. Here we present the results of the deepest VLA interferometric observation of Geminga performed so far.

\section{VLA Observations and Data Analysis}\label{vla-analysis}

Geminga was observed at 4.8 GHz (6 cm), with a bandwidth of 50 MHz with the VLA in D configuration (program AP468). The observation was performed on 2004 July 24 for a total of 6.0 on-source hours, with the antennas pointed at $\rm R.A.(J2000)  =06^h33^m54\fs02$, $\rm Dec.(J2000)=17\degr46\arcmin11\farcs5$. Calibration and imaging were performed with the NRAO Astronomical Image Processing System (\textsc{aips}) software package.\footnote{See http://www.aips.nrao.edu/.} The flux-density scale was calibrated by observing 0137+331 (3C48). The source 0625+146 was observed at intervals of 30 minutes and used as phase calibrator. The surface brightness image was produced following the standard procedures: Fourier-Transform, Clean, and Restore implemented in the  \textsc{aips} task \textsc{imagr}. We averaged the 2 IFs  together in the gridding process under \textsc{imagr}. Self-calibration was applied to remove residual phase variations.

\begin{figure}
\resizebox{\hsize}{!}{\includegraphics[angle=00,width=\textwidth]{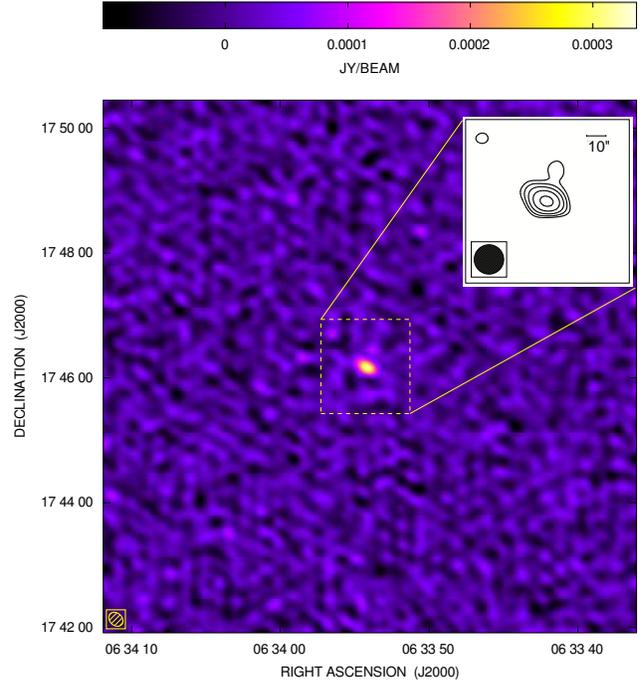}}
\caption{\label{f1}
Geminga field ($\sim$$8\arcmin \times 8\arcmin$) at 4.8 GHz resulting from the 6-hours VLA observation performed in D configuration. The beam FWHM is $15\farcs2\times 11\farcs8$ and the image sensitivity  is 25 $\mu$Jy beam$^{-1}$ (1$\sigma$). A zoom of the image, restored with a circular beam of $15.3\arcsec \times 15.3\arcsec$, is inset in order to evidenciate the slightly elongated nature of the radio feature. The faintest drawn radio contour level corresponds to 75 $\mu$Jy beam$^{-1}$ and the other levels are spaced by a factor of $\sqrt{2}$.}
\end{figure}

In Fig.~\ref{f1} we show a $\sim$8$\arcmin$-side field of the radio continuum emission around Geminga. The image  has a FWHM beam of $15\farcs2\times11\farcs8$ (position angle, $PA=46.5\degr$) and a noise  level of 25 $\mu$Jy beam$^{-1}$ (1$\sigma$). A radio feature, with a peak of brightness at $>$10$\sigma$ significance level, located at $\rm R.A.(J2000)=06^h33^m54\fs22~(\pm0\fs04)$, $\rm Dec.(J2000)=17\degr46\arcmin11\farcs22~(\pm0\farcs42)$ (through \textsc{aips} task \textsc{jmfit}), has been detected. The absolute astrometric accuracy of VLA observations is linked to the uncertainty in the phase calibrator position, which is better than $0\farcs15$ in our case. However, the VLA accuracy can be limited by a number of effects, such as the atmospheric phase stability.\footnote{For more details, see the VLA observation status summary at http://www.vla.nrao.edu/astro/guides/vlas/current/.} For this reason, we adopted for the positioning uncertainty a more conservative figure of $1\farcs8$, which was estimated as the synthesised beam size divided by the signal to noise ratio (see, e.g., \citealt{hagiwara01}).

The observed brightness distribution is more extended than that expected for a point source: it  is slightly elongated and fits with a single 2-dimensional Gaussian model with a FWHM major axis of $18\farcs6 \pm 1\farcs6$, a minor axis of  $11\farcs1 \pm 1\farcs0$, and $PA=(64.3\pm6.6)\degr$, incompatible with beam parameters (the fit with the point-spread function yields a poor reduced $\chi^2$ of $\sim$2.5 for 5 degrees of freedom\footnote{ We checked our capability to distinguish point-like from slightly elongated sources, such as the detected feature, by performing detailed simulations in which we artificially injected point sources in the raw VLA data.}). The deconvolved size along the major axis (approximately the direction of Geminga proper motion) is $11\arcsec\pm3\arcsec$, while it is unresolved in the transverse direction. The flux density of the source at 4.8 GHz is $(0.37\pm0.05)$ mJy being calculated by integrating the total intensity surface brightness down to the noise level. No evidence of source variability was found down to $\sim$1.5--2 hours, which is the minimum time-scale that can be probed in our data. At the distance of Geminga, the corresponding radio power is $S_{\mathrm{4.8\,GHz}}\simeq2.7\times10^{9}$ W Hz$^{-1}$.

To investigate the presence of such source in other VLA observations, we analyzed an archival dataset at 1.5 GHz in B configuration (program AK147). In this case we obtained an image with a FWHM beam of $4\farcs10\times3\farcs98$ and noise level of 22 $\mu$Jy beam$^{-1}$ (1$\sigma$). We did not found any source in correspondence to the 4.8 GHz detection. \citet{giacani05} reported on a study of the H\textsc{i} (21-cm line) interstellar gas distribution on a $40\arcmin\times40\arcmin$ field around Geminga based on VLA and Effelsberg radio telescope data. Based on the line-free channels of the 1.4 GHz cube, they produced a radio continuum image with 24\arcsec\ angular resolution, confirming the nondetection of any source at this frequency, either point-like or extended, down to a noise level of 0.14 mJy beam$^{-1}$ (1$\sigma$). 

Hence, the presently available data do not allow one to constrain the thermal or non-thermal nature of the emission, but, if the radio source is steady, we can set an upper limit on the spectral index\footnote{$S(\nu)\propto \nu^{- \alpha}$, where $\alpha$ is the spectral index.}  of $\rm\alpha^{4.8\,GHz}_{1.4\,GHz}<0.1$ (this value has been calculated by considering that  at 1.4 GHz the total flux density is below 3$\rm\sigma_{1.4\,GHz}=0.42$ mJy). We note however that if the radio source is variable, this limit is unreliable,  since the reported observations are not simultaneous.

\begin{figure}
\resizebox{\hsize}{!}{\includegraphics[angle=00,width=\textwidth]{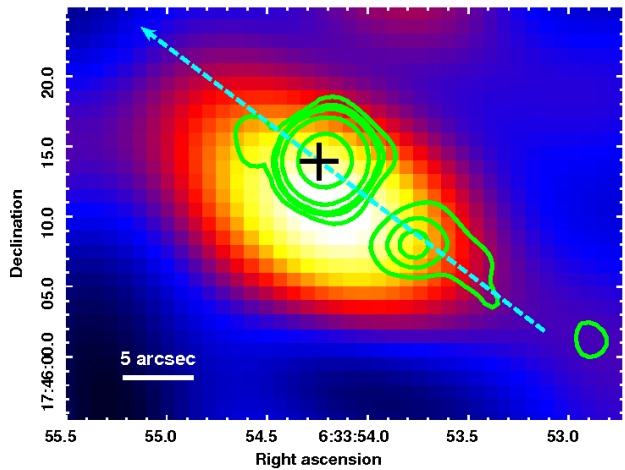}}
\caption{\label{f2} Detail of the radio feature shown in Fig.1. Black cross marks Geminga pulsar position at the epoch of VLA observation (July 2004) and the dashed arrow indicates its proper-motion direction. Green contour levels represent the X-ray emission (pulsar and PWN) detected by \emph{Chandra} in February 2004 (see Section~\ref{vla-analysis}). Most of X-ray axial features \citep{pavlov10} are detected within $\sim$10\arcsec\ ahead and behind the radio peak emission.} 
\end{figure}

We computed the coordinates of the Geminga pulsar at the epoch of the reported VLA observation by projecting the accurate optical absolute position by \citet{caraveo98} using the proper motion measurement by \citet{faherty07}: $\rm RA=06^{h}33^{m}54\fs24$, $\rm Dec.=+17\degr46'13\farcs9$ with errors $<$0\farcs1. The pulsar is within the reported diffuse feature, marginally displaced by $2.7\pm1.8$ arcsec from the radio peak emission and in association with a radio brightness of 0.26 mJy beam$^{-1}$ ($\sim$80\% of the peak flux; Fig.~\ref{f2}).The probability of random association between the radio feature and the Geminga PWN region (approximately FWHM beam size divided by the field of view) is $<$$10^{-3}$. We examined radio, optical and near-infrared catalogs looking for possible background objects of different nature (e.g. an AGN jet), but no counterpart compatible with the observed structure was found.

In order to compare the radio and X-ray morphologies of the structures associated to Geminga, we retrieved a public  \emph{Chandra} ACIS-S imaging exposure of the field performed close to the VLA observation (Obs. ID 4674; \citealt{dcm06,pavlov06}). The $\sim$20-ks-long observation was carried out on 2004 February 07 (168 days before the VLA observation) in `Timed Exposure' mode using a 1/8 subarray. Geminga was positioned in the back-illuminated ACIS-S3 chip, sensitive to photons between 0.2 and 10 keV. The data were processed using the \emph{Chandra} Interactive Analysis of Observation software (\textsc{ciao}, version 4.2) and we employed the  \textsc{caldb} 4.3 calibration files. Standard screening criteria\footnote{See \mbox{http://asc.harvard.edu/ciao/threads/index.html}.} were applied to generate a 0.5--8 keV image of the field of Geminga. Using the \emph{Chandra} aspect tool,\footnote{See \mbox{http://cxc.harvard.edu/cal/ASPECT/fix\_offset/fix\_offset.cgi}.} we found a small offset in the astrometry ($\Delta\mathrm{RA}=0\farcs01$, $\Delta\mathrm{Dec.}=0\farcs1$). After fixing the offset, in order to assess the absolute astrometry of the \emph{Chandra} data set, we computed the position of Geminga with the \textsc{wavedetect} task. We localized it at $\sim$$0\farcs3$ from the position expected on the basis of the source absolute optical position and proper motion, well within the typical ACIS-S localization accuracy.\footnote{See http://cxc.harvard.edu/cal/ASPECT/celmon.} The X-ray contours obtained from the 2004 \emph{Chandra} observation and related to the pulsar and to the axial PWN tail (see  also \citealt{dcm06,pavlov06}) are shown in Fig.~\ref{f2}.
The deconvolved length of the radio feature ($\sim$10\arcsec) is smaller than the whole time-integrated (2004 and 2007 \emph{Chandra} combined data) X-ray tail, but larger than the substructures (typically $\sim$5\arcsec-long `blobs') detected by \emph{Chandra} in individual observations \citep{pavlov10}.

\section{Discussion}\label{disc}

A VLA observation at 4.8 GHz unveiled a $\sim$$10\arcsec$-long radio feature with flux of $\sim$0.4 mJy associated with the Geminga pulsar and its PWN.
Significant contamination from Geminga persistent pulsed radio emission is very unlikely: given the tight upper limits on pulsed flux density at low frequency (e.g. $<0$.05 mJy at 0.4 GHz; \citealt{burderi99}), the contribution of pulsed emission  to the total observed flux at 4.8 GHz  would be $<$0.1\% for the mean spectral index of radio pulsars ($<$$\alpha$$>= 1.8\pm0.2$; \citealt{maron00}) and could at most reach $\sim$15\% even in the extreme case of a flat spectrum. The occurrence of transient pulsed (or continuum) radio emission in isolated NSs in not unheard-of, but it involves at our present knowledge pulsar classes showing a totally different phenomenology from that of the Geminga pulsar, like magnetars and rotating radio transients \citep{rea11,mclaughlin06short}. Furthermore, the emission models mentioned above, as well as any other kinds of magnetospheric emission originating very close to the NS surface (and hence appearing as point-like at the spatial resolution of the available observations), are disfavored by the evidence for a slightly elongated feature, together with the occurrence of a marginal offset between the NS position and the radio emission centroid ($\sim~2\farcs7$).

The latter two features and the tendency of the radio emission to distribute along the direction of the axial X-ray tail, naturally lead to an association with the Geminga PWN. \citet[][and references therein]{pavlov10} observed that the axial tail may result from the shocked pulsar wind collimated by the ram pressure (i.e. a bow-shock) or from a jet emanating along the pulsar's spin axis and perhaps aligned with the direction of motion. In the frame of the standard bow-shock models, radio PWNe are generally interpreted as arising from cooling of X-ray-emitting leptons \citep{kennel84}. The smaller size of the radio nebula of Geminga (with respect to the whole, time-integrated, $\sim$50\arcsec\ long X-ray tail) apparently contradicts this hypothesis, unless we associate the observed radio feature to the variable streaming `blobs' of $\sim$5\arcsec\ diameter, possibly constituting the X-ray axial tail as observed in 2004 \citep{pavlov10}.\footnote{Note that since presently available \emph{Chandra} and VLA observations are not simultaneous and the X-ray axial tail is partially hindered by the bright NS, we cannot in any case unequivocally link the observed radio feature to a specific substructure of the tail.} Furthermore, synchrotron models explaining the X-ray tails as pulsar wind shocked by the ram pressure due to the NS's supersonic motion typically require relatively low magnetic fields ($\sim$10--100 $\mu$G; \citealt{caraveo03}; \citealt{pavlov06}). With such low magnetic fields, the radio emission arising from the cooling of the X-ray emitting electrons, would be placed well outside the VLA field of view due to the large proper motion of Geminga: e.g. more than 0.5 deg off the position of the X-ray emission for a magnetic field of 100 $\mu$G.\footnote{See also e.g. \citealt{wang01} and \citealt{gaensler04} for examples of parameters of typical bow-shock PWNe simultaneously detected both in X-rays and radio.} In summary, ram-pressure collimated PWN models can be applied to our case only if additional processes, such as adiabatic expansion from hypothetical over-pressurized substructures of the tail (not observable at the present stage) provide a cooling power much higher than the synchrotron one and/or if radio emitting electrons are directly accelerated at the PWN shock, i.e. they are not originated from the cooling of high-energy electrons as in standard PWN models. In this respect, we note that the radio ``wisps" observed in the Crab PWN are produced by the same process as the higher-energy ones, although the mechanism by which this particle spectrum is generated is still not known (see e.g. \citealt{bietenholz06}).


Alternatively, the radio tail of Geminga could in principle be ascribed to a (magnetically-collimated) jet, in which the radio emission could directly result from radio-emitting electrons freshly-injected along the spin-axis at the jet formation site, or from internal shocks in which the accelerated X-ray-emitting leptons cool down to radio energies on the observed short space and time scales (see e.g. \citealt{benford84,komissarov04,meier03} for possible models of magneto-hydrodynamic jet produced by isolated pulsars). 

In the latter case, magnetic fields $\ga$0.05 G are required in order to avoid the occurrence of an additional (i.e other than synchrotron) cooling process. As already noticed for the case of the ram-pressure collimated PWN models, there is no evidence of such high fields in other PWNe. However, in the first place, large magnetic fields due to the NS dipole are in principle available at the pulsar light cylinder ($B_{\rm lc}\sim 10^3$ G for Geminga) and in its surroundings. In the second place, an observational bias may also be at work here. In fact, despite the growing number of jet-like structures around NSs resolved in X-rays \citep{kargaltsev08}, no `genuine' magneto-hydrodynamic radio jet around pulsar was detected so far (see e.g. \citealt{cohen83} for VLA mapping of numerous pulsar fields), possibly excepting PSR\,J2021+4026, another radio-quiet gamma-ray pulsar, located at the edge of the supernova remnant G\,78.2+2.1 \citep{trepl10}. Thus, it cannot be excluded that pulsar radio jets might be observable only in nearby and radio-quiet pulsars, not hindering their close environment by bright magnetospheric radio emission and having magnetic fields much stronger than those related to other PWN structures placed at larger distances from the NS surface. Interestingly enough, this hypothesis may be checked by a series of deep radio interferometric exposures. Scaling our result for Geminga, this survey could unveil jet-like structures in few other near ($\la$1 kpc) and energetic ($E_{\mathrm{rot}}\ga 10^{34}$ erg s$^{-1}$) radio-quiet pulsars, some of them recently discovered by the Fermi satellite \citep{abdo09}.

Finally, we note that a further possible explanation of the Geminga radio tail is the leakage of radio-emitting electrons (with Lorentz factor $\gamma\sim10$--100) directly from the open field lines of the magnetosphere. The maximum current through the magnetospheric accelerator (i.e., the Goldreich-Julian current; see e.g. \citealt{wang98} and references therein) and then the maximum particle flux escaping from Geminga open field lines can be estimated from $\dot{N}\simeq\Omega^2B_{\mathrm{surf}}R^3/2ec\simeq 5\times10^{31}$ s$^{-1}$ (where $\Omega$ is the pulsar angular velocity, and the NS's radius $R$ is assumed to be 10 km). The expected synchrotron power at 4.8 GHz, for a random distribution of pitch angles, can be roughly obtained from:
\begin{displaymath}
W^{\mathrm{synch}}_{\mathrm{radio}}\simeq \frac{4}{3}\sigma_T\gamma 
^2U_Bc\dot{N}\epsilon\tau_{\mathrm{synch}}\simeq 
10^{25}\gamma\epsilon~\rm erg~s^{-1}
\end{displaymath}
where $\tau_{\mathrm{syn}}$=$m_e$$c$/$(4\sigma_T\gamma U_B)$ is the synchrotron cooling time of the electrons, $U_B=B^2/(8\pi)$ is the magnetic field energy density and $\epsilon$ is the fraction (w.r.t. $\dot{N}$) of radio emitting particles. The observed luminosity of 10$^{26}$ erg s$^{-1}$ at 4.8 GHz would then be matched for $\gamma\epsilon\sim 10$, implying a magnetic field of $B\sim 20\epsilon^2$ G in the area of the emission. If this magnetic field is associated to the NS dipole, the emitting region would then be placed at a distance of $D_{\rm em}\sim 5\times 10^9\epsilon^{-2/3}$ cm from the NS surface. In summary, for $\epsilon=0.5$ (in turn implying $D_{\rm em}$ of the order of ten light cylinder radii, $\gamma\sim 20$, $B\sim 5$ G, and $\tau_{\mathrm{syn}}\sim 0.05$ yrs), this model would provide a cooling and possibly expanding diffuse radio feature protruding from the Geminga pulsar. If this model is correct, two geometrical constraints should be simultaneously satisfied by the Geminga pulsar: the radio pulsar beam not intersecting the line of sight and the leaking electrons producing a relatively narrow cone of radio emission roughly aligned with the projection of the pulsar proper motion. This would suggest that Geminga is a nearly aligned rotator, with the spin axis almost aligned with the proper motion and with the velocity vector close to the plane of the sky.

\section{Conclusions}

We have reported on the discovery of a radio emitting structure with a flux of $\sim$0.4 mJy at 4.8 GHz, associated with the Geminga pulsar and roughly aligned with the direction of the neutron star proper motion. 
The source appears slightly extended ($\sim$$10\arcsec$-long) but further deeper observations are needed to confirm the elongation unambiguously.
The nature of this emission is unconstrained by present data, but a combination of dedicated radio interferometric and X-ray observations\footnote{In particular, bulk flow velocity measurements in the radio band could assess the indication of motion/variability of X-ray blobs along the axial tail, suggestive of a jet-like interpretation, obtained by comparing \emph{Chandra} observations at different epochs \citep{pavlov10}.} is expected to shed light on the underlying physical process and likely unveil similar radio tails in other radio-quiet pulsars.

\section*{Acknowledgments}
VLA is operated by the  National Radio Astronomy Observatory (NRAO), a facility of the National Science Foundation (NSF) operated under cooperative agreement by Associated Universities, Inc. This research has also made use of data obtained from the \emph{Chandra} Data Archive and software provided by the \emph{Chandra} X-ray Center. P.E. acknowledges financial support from the Autonomous Region of Sardinia through a research grant under the program PO Sardegna FSE 2007--2013, L.R.  7/2007, ``Promoting scientific research and innovation technology in Sardinia''.  The authors are grateful to the anonymous referees, whose valuable comments led to substantial improvements in the paper.


\bsp

\label{lastpage}

\end{document}